\documentclass[aps,pre,twocolumn,superscriptaddress,showpacs,floatfix]{revtex4-1}
\usepackage[latin1]{inputenc}
\usepackage{graphicx}
\usepackage{amsmath}
\usepackage{amssymb}
\usepackage{pifont}
\usepackage{epstopdf}

\renewcommand{\Re}{\mathrm{Re}}
\newcommand{\Pe}{\mathrm{Pe}}

 \usepackage{color}

\begin{document}
\title{Advective superdiffusion in superhydrophobic microchannels}

 \author{Tatiana V. Nizkaya}
  \affiliation{A.N. Frumkin Institute of Physical Chemistry and
Electrochemistry, Russian Academy of Science, 31 Leninsky Prospect,
119071 Moscow, Russia}

 \author{Evgeny S. Asmolov}
\affiliation{A.N. Frumkin Institute of Physical Chemistry and
Electrochemistry, Russian Academy of Science, 31 Leninsky Prospect,
119071 Moscow, Russia}
\affiliation{Institute of Mechanics, M.V. Lomonosov Moscow State
University, 119991 Moscow, Russia}
 \author{Olga I. Vinogradova}
\email[Corresponding author: ]{oivinograd@yahoo.com}
 \affiliation{A.N. Frumkin Institute of Physical Chemistry and
   Electrochemistry, Russian Academy of Science, 31 Leninsky Prospect,
   119071 Moscow, Russia}
 \affiliation{Department of Physics, M.V. Lomonosov Moscow State
   University, 119991 Moscow, Russia}
 \affiliation{DWI - Leibniz Institute for Interactive Materials, Forckenbeckstr. 50, 52056 Aachen,
   Germany}

\date{\today}
\begin{abstract}
We consider pressure-driven flows in wide microchannels,
and discuss how a transverse shear, generated by misaligned superhydrophobic walls,  impacts cross-sectional spreading of Brownian particles. We show that such a transverse shear can induce an advective superdiffusion, which strongly enhances dispersion of particles compared to a normal diffusion, and that maximal cross-sectional spreading corresponds to a crossover between its subballistic and superballistic regimes. This allows us to argue that an advective superdiffusion can be used for boosting dispersion of particles at smaller Peclet numbers compared to known concepts of passive microfluidic mixing. This implies that our superdiffusion scenario allows one efficient mixing of much smaller particles or using much thinner microchannels than methods, which are currently being exploited.

\end{abstract}
\maketitle
\section{Introduction}
\label{sec:introduction}

Superhydrophobic (SH) textures in the Cassie
state, where the texture is filled with gas, have motivated numerous studies during the past decade~\cite{quere.d:2005,darmanin.t:2014}. Such surfaces  are important due to their superlubricating
potential~\cite{bocquet2007,Ybert2007,rothstein.jp:2010,vinogradova.oi:2012,feuillebois.f:2009}. The use of
highly anisotropic SH textures with generally tensorial effective hydrodynamic slip, $\mathbf
b_{\rm eff}$~\cite{stone2004,bazant2008tensorial,feuillebois.f:2009,harting.j:2012} (due to secondary
flows transverse to the direction of the applied pressure
gradient~\cite{mixer2010,vinogradova.oi:2010}), provides new possibilities for hydrodynamic flow
manipulation~\cite{ng.co:2010b,nizkaya2015flows,ng.co:2010,harting.j:2012}. Recent studies have employed transverse components of flow in SH channels to fractionate large non-Brownian microparticles~\cite{pimponi.d:2014,LabChip} or enhance their mixing~\cite{ou2007enhanced,jaimon2016numerical}. However, we are unaware of any previous work that has addressed the issue of diffusive transport of tiny Brownian particles by generating transverse flows in SH devices.

Diffusive transport controls diverse situations in biology and chemistry~\cite{frey.e:2005}, and its understanding is very important in many areas including such as nanoswimmers propulsion~\cite{wang.z:2014} or interpretation of modern nanovelocimetry experiments~\cite{vinogradova.oi:2009}. Dispersion of tiny Brownian particles in a cross-section of a microchannel with smooth walls  at low Reynolds number $\Re$, which is relevant to many applications, is difficult since the normal diffusion (characterized by the linear time dependence of the mean squared displacement, $\sigma^2 \propto t$) is slow  compared with the convection of particles along the microchannel. Our strategy here is to enhance such a dispersion by using advective diffusion, which can be induced by generating a transverse component of flow. Transverse flow generated by herringbone patterns in the Wenzel state (when liquid follows the topological variations of the surface) has been already successfully  used for a passive chaotic mixing of particles in a microchannel of thickness $H$ comparable to its width $W$ and at very large Peclet number, $\Pe$~\cite{ottino2004introduction,stroock2002chaotic,howell2008}. Here we suggest that dramatic improvement of a cross-sectional dispersion in a very wide channel, $W\gg H$, and at much smaller $\Pe$ (which is equivalent to significantly reduced particle sizes or channel thickness) could be achieved by inducing a superdiffusion, i.e. a situation, when $\sigma^2 \propto t^{\alpha}$ with $\alpha > 1$. Depending on the value of $\alpha$ one usually distinguishes between subbalistic ($1< \alpha < 2$), ballistic ($\alpha = 2$), and superballistic ($\alpha > 2$) regimes of superdiffusion~\cite{metzler2000}. The superdiffusion in a flow field has been studied by several
groups for various macroscopic systems. Subballistic regime has been reported for a random velocity fields~\cite{matheron1980,krapivsky2010}, and superballistic dispersion has been predicted for turbulent~\cite{shlesinger1987} and for linear shear~\cite{foister1980} flows,  and for a solute transport in a heterogeneous medium~\cite{dentz2003}.
Some  efforts have also gone into investigating a role of confinement in the emergence of superdiffusion~\cite{Benichou.o:2013}.
However, advective superdiffusion on microscales has never been predicted theoretically, nor has it been used for microfluidic applications.

The presence of an additional variable $H$ in the system implies that
diffusive behavior in a confined complex flow should be different than it would be in bulk liquid or near a single interface. Could various  superdiffusive regimes be induced in microchannels with realistic parameters of the flow? How will they differ from the bulk systems if induced? What are possible implications for microfluidic mixing? These questions still remain open, and we are unaware of any previous attempts describing answers to them.

  \begin{figure}[t]
  \includegraphics[width=1\columnwidth]{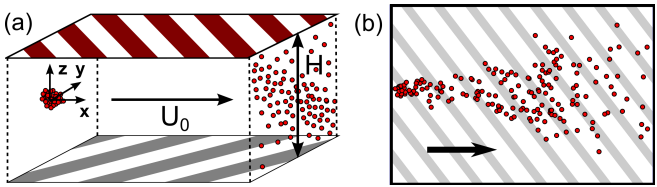}
  \caption{(a) Sketch of the superhydrophobic channel with identical, but misaligned striped textures at the walls. Main flow direction is from left to right.
(b) Top view of particle dispersion. }
  \label{fig:sketch1}
\end{figure}

In this paper we present a general strategy for inducing an advective superdiffusion in microchannels of a high aspect ratio, $W/H\gg 1$, that can
be used for boosting dispersion of Brownian particles between streams of main (forward) Poiseuille
flows. To enhance the mixing (homogenization) of particles over the cross section of the channel we use secondary (transverse) shear flows generated in microchannels decorated by crossed identical SH stripes~\cite{nizkaya2015flows} as sketched in Fig.~\ref{fig:sketch1}. We show that such a flow configuration allows one to induce various scenarios of a superdiffusion, and argue that a crossover between subballistic and superballistic regimes provides large transverse dispersion of particles, which would be impossible in standard microchannels with smooth homogeneous walls or in devices, which are currently widely used as microfluidic mixers.

\section{Scaling theory}

We first present the scaling approach which we have developed to evaluate hydrodynamic dispersion in a Poiseuille flow with superimposed uniform transverse shear:
\begin{equation}
\begin{array}{ll}
U_x=U_{sx}+U_0\left( 1-4(z/H)^{2}\right),\\%
U_y=2U_{sy}z/H,
\end{array}
\label{velocity}
\end{equation}
where $U_0=-H^2\nabla P/(8\mu)$ is the maximal velocity of a flow generated by pressure gradient $\nabla P$  with no-slip walls, $\mu$ is the dynamic viscosity, $U_{sx}$ and $U_{sy}$ are the (positive definite) averaged forward and transverse slip velocities at channel walls located at $z=\pm H/2$. We note that the transverse shear rate is equal to $2U_{sy}/H$.

Brownian particles are injected from a point source located at $(x,y,z)=(0,0,0)$ and then advected by the flow satisfying Eq.(\ref{velocity}). The particle flux across channel walls is equal to zero, we neglect their inertia, and focus on the diffusive regime.

Since both $U_x$ and $U_y$ depend only on $z$, particle distribution in $z$-direction will be governed by normal Brownian diffusion  with zero average displacement. For an unbounded space we have:
$\langle z\rangle=0,\;\sigma_z^2=\langle (z-\langle z\rangle)^2\rangle= 2Dt$, where $\langle.\rangle$ denotes averaging over the ensemble of particles, and $D$ is the diffusion coefficient. In our case diffusion is constrained by channel walls, so that some time later particles become uniformly distributed between them: \begin{equation}\begin{array}{ll}
\sigma_z^2=2Dt,& t\ll t_d,\\
\sigma_z^2=H^2/12 ,& t\gg t_d
\label{sigmaZ}
\end{array}
\end{equation}
Here we have defined the diffusion time scale, $t_d=H^2/(2D)$, as a typical time for a single particle to cross the channel in $z$-direction.

Particle dispersion in $y$-direction reflects an interplay between diffusion and transverse shear rate. Depending on $t$ different scenarios of the particle spreading may occur.
In the short time regime, i.e. for $t\ll t_d$, our shear flow could be treated as unbounded, since the spreading of particles is still unaffected by confinement. By substituting the expression for a transverse shear rate into a solution for a mean square displacement of Brownian particles in an unbounded linear shear~\cite{foister1980} we obtain
\begin{equation}\label{eq:sigma_y}
\sigma_y^2=2Dt\left(1+\dfrac{1}{3}\left(\dfrac{2U_{sy}t}{H}\right)^2\right),\;	t\ll t_d.
\end{equation}
This expression defines a second time scale $t_s=H/(2U_{sy})$, which is associated to a transverse shear.

We note that depending on the value of the Peclet number, $\Pe=U_0 H/D$, the ratio $t_d/t_s=U_{sy}\Pe/U_0$ can vary in a large interval. Two limits can now be discussed depending on the ratio $t_d/t_s$. When $t_d/t_s \ll 1$, which is equivalent to $\Pe \ll U_0/U_{sy}$,   the normal Brownian diffusion of particles provides their efficient spreading in the channel since $D$ is large. However, if $t_d/t_s \gg 1$, which corresponds to $\Pe \gg U_0/U_{sy}$ (or small $D$), the normal diffusion is slow. Therefore, below we discuss this larger $\Pe$ limit in more detail. We note that in this situation if $t\ll t_s$, the mean square displacement of particles scales as $\sigma^2_y\propto D t$, indicating a normal Brownian diffusion. However, for $ t \gg t_s$, we deduce from Eq.(\ref{eq:sigma_y}) $\sigma^2_y \propto  D t^{3} U^2_{sy} H^{-2}  $, which suggests a superdiffusion of particles in a superballistic regime.

In the long-time regime, $t\gg t_d$, multiple rebounds of particles from the channel walls  should inevitably lead to a random variation of the transverse velocity even in our directed shear flow. This situation is similar to considered in prior work~\cite{matheron1980,krapivsky2010} on  Brownian particles in a (bulk) random velocity field, which predicted $\sigma^2_y \propto t^{3/2}$.
The characteristic velocity and length are determined by $U_{sy}$ and $H$, so that dimensional analysis immediately leads to
\begin{equation}
\sigma^2_y \propto
D^{-1/2}U^2_{sy} H t^{3/2},\;	t\gg t_d.
\end{equation}

We now summarize the different scaling expressions for $\sigma^2_y$, which determine several diffusion-advection regimes when $\Pe \gg U_0/U_{sy}$, and turn to dimensionless parameters:
\begin{equation}
\sigma^2_y /H^2 \propto \left\{\begin{array}{ll} t/t_d, & t\ll t_s\\
(U_{sy}/U_0)^2 \Pe^2 (t/t_d)^{3}, & t_s\ll t\ll t_d\\
(U_{sy}/U_0)^2 \Pe^2 (t/t_d)^{3/2}, & t_d\ll t,\end{array}\right.
\label{scaling_t}
\end{equation}

Eqs.(5) include the ratio $U_{sy}/U_{0}$, which depends on the superhydrophobic texture topology only. We focus here on microfluidic applications, and therefore it is not the time, but the channel length $x = \lambda H$ serves as a main independent parameter of the problem. So we have to reformulate Eq.(\ref{scaling_t}) in terms of $\lambda$. A time required for particles to migrate along the channel is $t= \lambda H /U_m$, where $U_m$ is a mean forward flow velocity in the locus of the assembly of particles. At $t\ll t_d$ it is equal to the velocity at the midplane of the channel, $U_m=U_0$, but at $t\geq t_d$ this will be the mean forward velocity in the channel, $U_m=2U_0/3$.
Note that in both cases the relationship between $\lambda H$ and $t$ is linear, $t/t_d\propto \lambda/\Pe$. Therefore, Eq.(\ref{scaling_t}) can be rewritten  as
\begin{equation}
\sigma^2_y /{H^2} \propto\left\{\begin{array}{ll}\Pe^{-1} \lambda, & \lambda \ll U_{0}/U_{sy},\\
\left(U_{sy}/{U_0}\right)^2 \Pe^{-1} \lambda^{3}, & U_{0}/U_{sy}\ll \lambda \ll \Pe,\\
\left( U_{sy}/{U_0} \right)^2 \Pe^{1/2}\lambda^{3/2}, & \lambda\gg\Pe.\end{array}\right.\label{scaling_x}
\end{equation}
In the other limit of $\Pe \ll U_0/U_{sy}$, a normal diffusion is expected as discussed above, so that in this case we should also get $\sigma^2_y /{H^2} \propto \Pe^{-1} \lambda$.
For a given
diffusion coefficient we obtain for the superballistic regime $\sigma
_{y}^{2}\propto U_{0}^{2}t^{3}\propto U_{0}^{-1}$ (when the flow is too fast
the migration time is too small) while for the subballistic regime we get $\sigma _{y}^{2}\propto U_{0}^{2}t^{3/2}\propto U_{0}^{1/2}.$ These two scalings imply the existence of optimum $U_{0}$ and corresponding $\Pe_{\max }.$

 \begin{figure}[h!]
  \includegraphics[width=0.75\columnwidth]{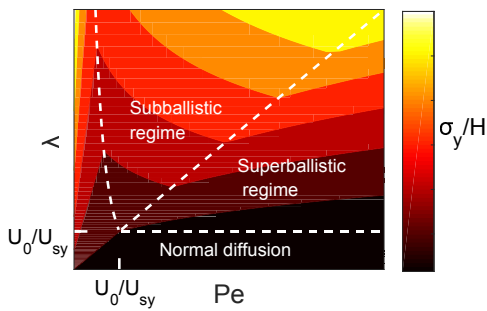}
  \caption{Schematic representation of various diffusion regimes in a microchannel. The colorbar values ascend from bottom to top.}
  \label{fig:Diagram}
\end{figure}

To illustrate this effect it is useful to divide the $(\lambda, \Pe)$ space into three regions of normal, subballistic and superballistic diffusion, where the above scaling expressions for $\sigma^2_y /{H^2}$ approximately hold. Such a diagram is plotted in Fig.~\ref{fig:Diagram}. The crossover loci here simply indicate that limiting solutions for $\sigma^2_y /{H^2}$ given by Eqs.(\ref{scaling_x}) coincide for different regimes of diffusion. Apart from
the curve $\lambda = \Pe^{3/2} (U_{0}/U_{sy})^2$ (separating the normal and superballistic diffusion), other crossover loci, $\lambda = \Pe$ (separating the subballistic and superballistic regions) and $\lambda = U_{0}/U_{sy}$ (between the normal and superballistic diffusion) are straight lines. Of course, in reality at those curves, the limiting solutions for $\sigma^2_y /{H^2}$ crossover
smoothly from one diffusion regime to another. We can now conclude that when $\lambda$ is below $U_{0}/U_{sy}$ only normal diffusion is expected. In other words, advective superdiffusion cannot be generated without large slip at SH walls. When $\lambda$ is above $U_{0}/U_{sy}$ three regimes can be attained depending on the value of $\Pe$. A very small Peclet number will lead to a normal diffusion, but at larger $\Pe$ one can induce a subballistic and at very large $\Pe$ - a superballistic regime. Fig.~\ref{fig:Diagram} also immediately shows that the maximal spreading is attained at the crossover between subballistic and superballistic regimes.
This means that it happens when the time required for particles to migrate forward to a given cross-section and to diffuse to the channel walls are comparable.

\section{Simulation method}

We model a  motion of Brownian particles, i.e. a situation when inertia is neglected and Langevin equations are reduced to first order

\begin{equation}
\dot{\mathbf{x}}=\mathbf{u}(\mathbf{x})+ \mathbf{r}(t),
\label{a1}
\end{equation}
where $\mathbf{x}$ is a particle position, $\mathbf{u}(\mathbf{x})$ is the velocity field of a fluid and $\mathbf{r}(t)$ is a random velocity component with a correlation time much smaller than other time scales in the system. To discretize the equation we introduce a time grid $\{t_k\}$ with the step $\Delta t$ and keep the random component  $\mathbf{r}_k=(r_{x}, r_{y},r_z)$ constant over the time step $[t_k,t_{k}+\Delta t]$ with random variables $r_{x,y,z}$ taken from the Gaussian distribution with zero average and dispersion $u_r$.

To validate that we model a dispersion of particles correctly we first measure their diffusion coefficient. We integrate the  equation of particle motion for an ensemble of $500$ particles released at $\mathbf x_0=(0,0,0)$ in a velocity field involving a uniform $\mathbf u= (U_0, 0, 0)$ and a random components $\mathbf{r}_k=(0, r_{y},r_z)$, and then measure the dispersion of particle positions $\sigma_{y,z}(t)$ at $t\leq T$. We then fit the dispersion curves using the standard scaling $\sigma_{y,z}^2(t)=2D_{y,z}t$. The diffusion coefficient is the same in all directions, $D_{z}=D_{y}=D$, and depends on the time step $\Delta t$ and the dispersion $u_r$ as $D=D^*u_{r}^2/(2\Delta t),$
where $D^*$ is a renormalization coefficient. We have computed the values of $D^*$ using $U_0=1$, $T=10$, several $u_r$ in the range from $0.05$ to $0.5$, and $\Delta t$ varying from $0.01$ to $0.1$. In all the cases, the simulations give $D^*=1$, which confirms that the scaling holds in the whole range of our parameters. 

The velocity field in the SH channel is calculated using the solution of Stokes equations valid at $H/L\simeq 1$ \cite{nizkaya2015flows}:

\begin{equation}
\begin{array}{ll}
\displaystyle\mathbf{u}=\left\langle \mathbf{u}\right\rangle +\mathbf{u}_{1}+\mathbf{u}_{2}.\\
\label{u}
\end{array}
\end{equation}
Here  $\left\langle \mathbf{u}\right\rangle=(U_x,U_y,0)$ is the averaged flow profile defined by Eq.(\ref{velocity}) and $\mathbf{u}_{1},$ $\mathbf{u}_{2}$ are perturbations with zero mean over the cell volume due to heterogeneous slippage at the
lower and upper walls, respectively. The perturbation fields $\mathbf u_{1}$ and $\mathbf u_{2}$ are obtained using Fourier series with $50$ harmonics \cite{nizkaya2013flow}.

In simulations of superdiffusive regimes we also use an ensemble of $500$ particles. To calculate the contribution of the fluid velocity field, $\mathbf{u}(\mathbf{x})$, the equation of motion, Eq.(\ref{a1}), is solved using 4-th order Runge-Kutta method with the time step $\Delta t=0.01 L/U_0$. A random component $u_r=\sqrt{2 D \Delta t}$ is constant over the time step. In these simulations bounce-back boundary conditions are applied at the channel walls.

\section{Results and discussion}

In order to assess the validity of the above scaling approach we now model a situation when the transverse shear is created by SH walls~\cite{nizkaya2015flows}, as sketched in Fig.~\ref{fig:sketch1}.
Specifically, we consider a pressure-driven flow between two parallel SH surfaces separated by the distance
$H$, which are decorated with identical periodic stripes of a period $L$ and a fraction of the gas area $\phi$. We assume SH surfaces to be flat with no meniscus curvature, so that the gas area is characterized by a local slip length $b$ only, and solid area has no-slip. The lower and the upper wall textures are misaligned by an angle $\pi/2$, and we align the $x$-axis and the pressure gradient with this angle bisector.
  The Reynolds number $\Re=\rho U_0H/\mu$, where $\rho$ is the fluid density, is considered to be small, i.e. $\Re\ll 1$, so that the flow satisfies  the Stokes equations.

\begin{figure}[h!]
\includegraphics[width=0.99\columnwidth]{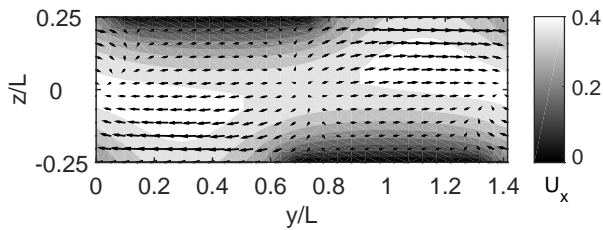}
\caption{Cross-section of the fluid velocity field $(u_y, u_z)$ at $\phi =0.5$ and $\lambda=0.5$. Colorbar shows the forward velocity $U_x$.}
   \label{fig:VelField}
\end{figure}

The typical velocity field in such a channel has been calculated following the method described before~\cite{nizkaya2015flows}, and its typical cross-section is shown in Fig.~\ref{fig:VelField}. Fig.~\ref{fig:VelField} illustrates that the transverse velocity is strongly inhomogeneous. The inverse average transverse velocity, $U_0/U_{sy}$, which controls the hydrodynamic dispersion, can be obtained by averaging the 3D velocity field $\mathbf{u}(x,y,z)$ over the periodic cell in $x,y-$plane. If $H = O(L)$ or larger, it can be evaluated by using a simple expression $ U_0/U_{sy}\simeq (1+2\beta_+)/(4\beta_-)$, where $\beta_{\pm}=(b_{\text{eff}}^\|\pm b_{\text{eff}}^\perp)/(2H)$ with $b_{\text{eff}}^{\|,\perp}$ the eigenvalues of the slip length tensor, $\mathbf b_{\rm eff}$, for a channel of a finite $H/L$ with one SH and one no-slip hydrophilic walls~\cite{nizkaya2015flows}. These eigenvalues have been calculated before~\cite{harting.j:2012}, and they depend on $b,$ $L/H,$ and $\phi$. In our simulations below we use $b=\infty$ since it maximizes the effective slip. We employ $L/H=2$ since this provides a significant transverse shear~\cite{nizkaya2015flows}. Finally, we  consider several textures, with $\phi$ varying in the interval from $\phi=0.25$ to $0.9$, which with prescribed parameters give a variation of $U_0/U_{sy}$ from $\simeq 11.5$ to $1.2$. With these values at moderate Peclet numbers one can expect various regimes of superdiffusion at an appropriate value of $\lambda$.

\begin{figure}[h]
\includegraphics[width=0.99\columnwidth]{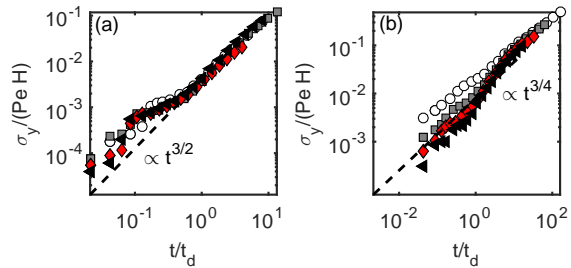}
\caption{Dispersion $\sigma_y/(H \Pe)$ as a function of time $t/t_d$ simulated for $\phi=0.5$. The Peclet numbers are (a) $\Pe=200,300,400,500$ (circles, squares, diamonds, triangles) and (b) $\Pe=10,50,75,100$ (circles, squares, diamonds, triangles).}
   \label{fig:scaling}
\end{figure}

We now inject a large number of Brownian particles in the channel, track their instantaneous positions and evaluate the dispersion $\sigma_y$ at a given time. We have first plotted in Fig.~\ref{fig:scaling} the simulation results for $\sigma_y/(\Pe H)$ as a function of $t/t_d$ obtained at $\phi=0.5$ and several $\Pe$. A general conclusion from this plot is that the above scaling predictions given by Eq.(\ref{scaling_t}) are in good agreement with simulation results. Thus, we see that all curves indeed overlap at long time. For relatively large Peclet numbers, $\Pe \geq 200$, simulation data confirm the superballistic scaling $(t/t_d)^{3/2}$, but at smaller Peclet numbers, i.e. $\Pe \leq 100$, our results fully validate the predicted subballistic scaling $(t/t_d)^{3/4}$.

\begin{figure}[h!]
\includegraphics[width=0.99\columnwidth]{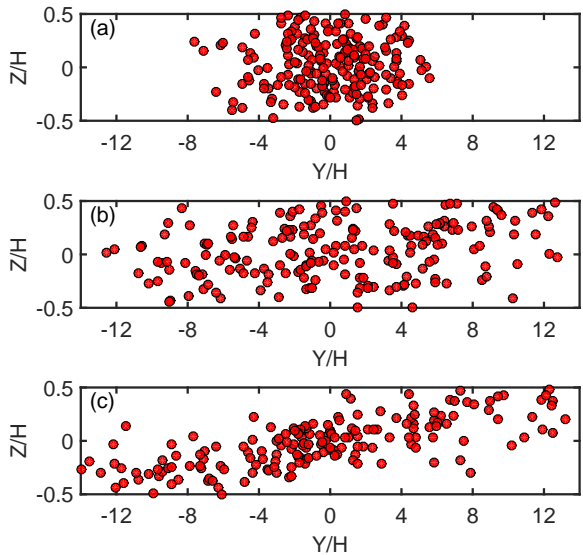}
\caption{Positions of individual particles at the cross-section $\lambda =50$ of a SH channel with $\phi =0.5$. The Peclet number is 10 (a),  50 (b), and 300 (c) providing different scenarios of advective superdiffusion.  }
 \label{fig:clouds}
\end{figure}

To examine the difference between different advective superdiffusion regimes we now vary $\Pe$ at fixed $\phi = 0.5$ and determine positions of individual particles at a given cross-section $\lambda =50$. For this gas area fraction $U_0/U_{sy}\simeq 7.4$, therefore, according to Fig.~\ref{fig:Diagram}, the values of $\Pe = 10, 50,$ and $300$  should lead to a subballistic regime of superdiffusion,  a crossover between subballistic and superballistic regimes, and a superballistic superdiffusion, correspondingly. The simulation results are shown in Fig.~\ref{fig:clouds}(a), (b), and (c). We see that a crossover between subbalistic and superballistic regimes does lead to a homogeneous distribution of particles. A subbalistic regime results in a rather homogeneous distribution of particles by the height of a channel, but $\sigma_y$ remains small, so that particles are still focussed near the midplane of the channel, $y=0$. In contrast, in the superballistic regime the particle spreading in the $y-$direction is large, but due to small $\sigma_z$ the distribution of particles in the cross-section is highly inhomogeneous.

 \begin{figure}[h]
\includegraphics[width=0.99\columnwidth]{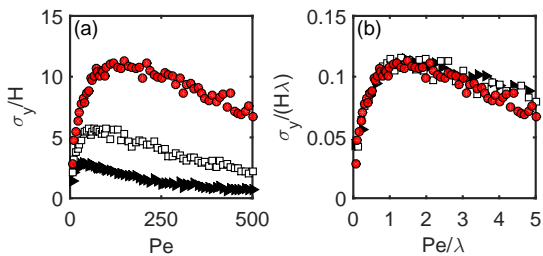}
\caption{(a) Transverse particle dispersion $\sigma_y / H$ as a function of $\Pe$ calculated with $\phi = 0.75$ at $\lambda =25$ (squares), $50$ (triangles), and $100$ (circles); (b) The same data plotted in scaled by $\lambda$ coordinates.} \label{fig:optimPe}
\end{figure}

Finally, we explore in more details a situation of a maximal transverse hydrodynamic dispersion, $\sigma_{\mathrm{max}}/H$, which occurs when the scaling law
\begin{equation}\label{pela}
\Pe_{\rm max} \propto \lambda,
\end{equation}
is valid. Fig.\ref{fig:optimPe}(a) shows $\sigma_y/H$, vs. $\Pe$, calculated at fixed $\phi=0.75$ and several $\lambda$. It can be seen that $\sigma_y/H$ increases with $\lambda$, and that for a given $\lambda$ there indeed exists a Peclet number, $\Pe_{\mathrm{max}}$, which maximize the dispersion. Note that the induced by superdiffusion transverse dispersion is large, already at moderate $\lambda = 50$ it could be several times larger than the channel thickness, of course, provided $\Pe$ is optimal. The scaling low, Eq.(\ref{pela}), predicts $\Pe_{\mathrm{max}}$ is growing linearly with $\lambda$. This is indeed the tendency shown by the simulation results. We now reproduce the data set from Fig.~\ref{fig:optimPe}(a) in Fig.~\ref{fig:optimPe}(b), but scale both coordinates by $\lambda$. Remarkably, and in agreement with our scaling analysis, simulation data obtained for several $\lambda$ do collapse into a single curve. This plot allows us to obtain a scaling prefactor in Eq.(\ref{pela}), which for a given $\phi=0.75$ is found to be $\simeq 1$ (see Appendix~\ref{A1}).

 \begin{figure}[h!]
\includegraphics[width=0.99\columnwidth]{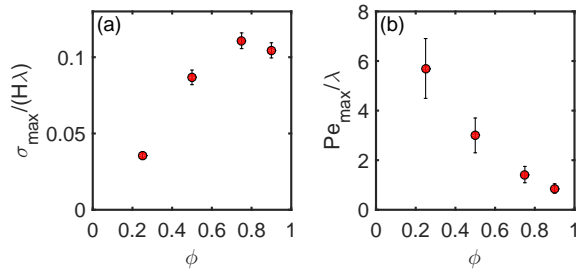}
\caption{Rescaled maximal dispersion, $\sigma_{\mathrm{max}}/(H \lambda)$, (a) and the Peclet number, $\Pe_{\mathrm{max}} / \lambda$, (b) as a function of the gas area fraction, $\phi$.} \label{fig:rescaledPe}
\end{figure}

Similar curves,  $\sigma_y/(H \lambda)$ vs. $\Pe / \lambda$, have been calculated for several gas area fractions, and we have again found that at a given $\phi$ they nearly coincide (see Appendix~\ref{A1}). We have then obtained from these simulation data the values of $\sigma_{\mathrm{max}}/(H \lambda)$ and $\Pe_{\mathrm{max}}/\lambda$ (which gives us exactly the scaling prefactor), and the results are plotted in Fig.\ref{fig:rescaledPe} as a function of $\phi$. A first conclusion emerging from this plot is that $\phi$ is one of the key parameters determining the maximal value of a transverse dispersion, $\sigma_{\mathrm{max}}/(H \lambda)$. Fig.~\ref{fig:rescaledPe}(a) shows that in the low $\phi$ the transverse hydrodynamic dispersion is very small. It grows with the gas area fraction and reaches the maximum at $\phi\simeq 0.75$  (see Appendix~\ref{A2} for interpretation of this result). We also note that the scaling prefactor in Eq.(\ref{pela}) decays with $\phi$ as seen in Fig.~\ref{fig:rescaledPe}(b).

Altogether the above simulation results do confirm our simple scaling lows. We can therefore  conclude that these expressions provide us with a correct picture of the superdiffusive behavior of particles in the flow, even thought they overlook many details.

\section{Final remarks}

In conclusion, we believe we have provided a satisfactory answer to several questions posed at the beginning of this paper. We have shown that by using wide microchannels with misaligned striped SH walls it is possible to induce an advective superdiffusion of Brownian particles, which could not be achieved a standard microfluidic devices with a smooth walls or in devices, which are currently used to enhance mixing at the microscale. We have developed scaling laws for  regimes of advective superdiffusion in such a microchannel,
providing explicit expressions for mean square displacement of particles as a function of channel thickness and length, Peclet number, and slip velocity at the walls. These scaling results have been validated by means of computer simulations. It has also been shown that the advective superdiffusion could be used to efficiently mix Brownian particles, which is important in a variety of applications.

Certain aspects of our work warrant further comments. A striking conclusion from our work is that the surface textures which optimize $\sigma_{\mathrm{max}}/(H \lambda)$ differ from those optimizing effective (forward) slip. It is well
known that the effective slip of SH surfaces is maximized by increasing the gas-liquid area fraction~\cite{feuillebois.f:2009,Ybert2007,harting.j:2012}. In contrast,
we have shown that dispersion of Brownian particles in SH microchannels is
maximized by stripes with a smaller gas fraction, $\phi\simeq 0.75$. In this situation,
the effective slip is relatively small, and yet the dispersion of particles is very strong.

We should also like to stress that in a bulk or near a single interface the optimal spreading of Brownian particles would obviously occur in a superballistic regime.
Our work has shown that that contrary to a bulk situation maximal cross-sectional spreading at a finite $H$ corresponds to a crossover between subballistic and superballistic regimes of a superdiffusion. Therefore, one can conclude that a superdiffusive behaviour of particles in a confined flow is indeed very different from expected for unbounded systems.

Prior work on passive micromixing~\cite{ou2007enhanced,stroock2002chaotic} has exploited microchannels of $W/H =O(1)$ and very large Peclet numbers, $\Pe=10^3-10^6$. This implies that previous methods have been designed to efficiently mix particles of a micron size or slightly smaller. We have addressed a different flow configuration of $W/H \gg 1$, and have argued that an advective superdiffusion can be used for boosting dispersion of particles at much smaller $\Pe$ compared to known concepts of passive microfluidic mixing. Our optimal $\Pe$ has been found to be of the order of $100$ and smaller. This implies that at the same velocities of a mean flow we could provide mixing of particles of the size $10-50 \,{\rm nm}$ (i.e. of truly nanoparticles, including proteins, viruses, etc). Alternatively, our concept allows one mixing of particles of the size of $0.1-1 \, \mu{\rm m}$ at the same flow rate, $Q=U_{0}HW$, but in channels of much smaller $H$.

\begin{acknowledgments}
This research was partly supported by the Russian Foundation for Basic Research (grant  15-01-03069).
\end{acknowledgments}

\appendix

\section{Calculations of optimal Peclet number and dispersion}\label{A1}

 \begin{figure}[h!]
  \includegraphics[width=0.99\columnwidth]{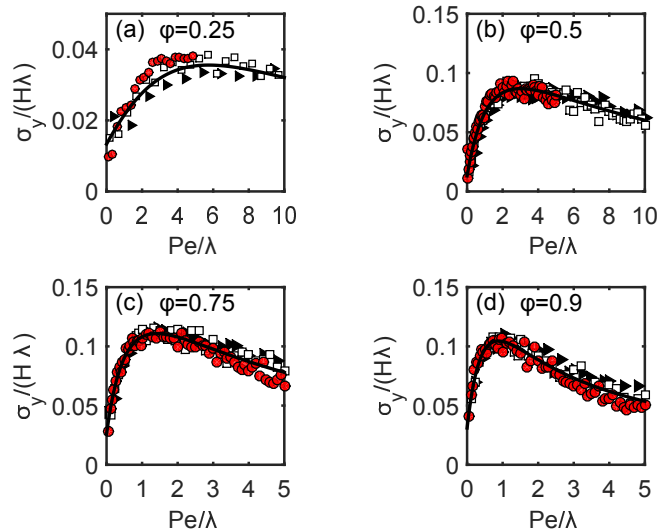}
  \caption{Particle dispersion $\sigma_y$ for $(a)\;\phi=0.25,\;(b)\;0.5,\;(c)\;0.75,\;(d)\;0.9$ and $\lambda=25,\;50,\;100$ (triangles, squares, circles). Solid curves show a fit to $ f_\sigma$, given by Eq.(\ref{eq:fit}).  }
  \label{fig:allphi}
\end{figure}

To estimate the optimal Peclet number and its dependence on texture parameters we run simulations for several $\lambda=25,50,100$ and several $\phi=0.25,\;0.5,\;0.75,\;0.9$. The $\sigma_y/(H\lambda)$ vs. $\Pe/\lambda$ curves are plotted in Fig.\ref{fig:allphi}. We see that at a given $\phi$ results for different $\lambda $ collapse into a single curve. To obtain the position of the maximum we fit these curves to a function
\begin{equation}\label{eq:fit}
 f_\sigma=\dfrac{p_1 x^2 + p_2 x + p_3 }{x^2 + p_4 x + p_5},
\end{equation}
where $p_i$, $i=1-5$, are the fitting coefficients. To find the best fit for $p_i$ we use the combined data set for all three values of $\lambda$.
The values $\Pe_\mathrm{max}/\lambda$ and $\sigma_\mathrm{max}/(H\lambda)$ are then calculated from the fitting function for each $\phi$.  The error bars in Fig.7 for the maximal dispersion $\Delta\sigma_{\mathrm{max}}$ are found by  estimating the root-mean-square deviation between the fit and the data. 
The error in $\Pe_{\mathrm{max}}$ is estimated by expanding the fit function around its maximum,

\begin{equation}
    \Delta\Pe\sim \sqrt{\Delta\sigma_{\mathrm{max}}/f''_\sigma(\Pe_{\mathrm{max}}/\lambda)}.
\end{equation}

\section{The effect of large forward slip on the dispersion}\label{A2}

The scaling Eq.(\ref{scaling_x}) has been obtained assuming that the slip velocity at the wall is small compared to that of the Poiseuille flow, $U_{sx},U_{sy} \ll U_{0}$, so that they do not include a forward slip $U_{sx}$. In this Appendix, we estimate the contribution of large forward slip, $U_{sx}$, to $\sigma _{y}$ and $\mathrm{Pe}_{\max }$. When $U_{sx},
U_{sy} \simeq U_{0}$ the mean forward flow velocity is
\begin{equation*}
U_{m}=\left\{
\begin{array}{ll}
U_{0}+U_{sx}, & t\ll t_{d}, \\
2U_{0}/3+U_{sx}, & t\gg t_{d}.%
\end{array}%
\right.
\end{equation*}

Since $t=\lambda H/U_{m},$  Eqs.(\ref{scaling_t}) and (\ref{scaling_x}) can be modified to give

\begin{equation*}
\dfrac{\sigma _{y}^{2}}{H^{2}}\propto \left\{
\begin{array}{ll}
\dfrac{\left( U_{sy}/{U_{0}}\right) ^{2}\lambda ^{3}}{\left( 1+U_{sx}/{U_{0}}\right) ^{3}\mathrm{Pe}}%
, \; U_{0}/U_{sy}\ll \lambda \ll \mathrm{Pe}, \\
\dfrac{\left( U_{sy}/{U_{0}}\right) ^{2}\mathrm{Pe}^{1/2}\lambda ^{3/2}}{\left( 2/3+U_{sx}/{U_{0}}\right)
^{3/2}}, \; \lambda \gg \mathrm{Pe}%
\end{array}%
\right.   \label{scaling_x2}
\end{equation*}
and
\begin{equation*}
\dfrac{\sigma _{y}^{2}}{\left( {H\lambda }\right) {^{2}}}\propto \left\{
\begin{array}{ll}
\dfrac{\left( U_{sy}/{U_{0}}\right) ^{2}}{\left( 1+U_{sx}/{U_{0}}\right) ^{3}}%
\mathrm{Pe/}\lambda , \; U_{0}/U_{sy}\ll \lambda \ll \mathrm{Pe}, \\
\dfrac{\left( U_{sy}/{U_{0}}\right) ^{2}}{\left( 2/3+U_{sx}/{U_{0}}\right)
^{3/2}}\left( \mathrm{Pe/}\lambda \right) ^{1/2}, \; \lambda \gg \mathrm{Pe}.%
\end{array}%
\right.   \label{scaling_x3}
\end{equation*}%

When $\phi
\to 1,$ both $U_{sy}$ and $U_{sx}$ become large, and the dispersion for the first regime decays with $\phi$ since $%
\sigma _{y}^{2}/{H^{2}}\propto U_{0}U_{sy}^{2}U_{sx}^{-3}.$ However, it grows
for the second regime, $\sigma _{y}^{2}/{H^{2}}\propto
U_{0}^{1/2}U_{sy}^{2}U_{sx}^{-3/2}.$
The values of $\mathrm{Pe}_{\max }$ and $\sigma _{\max }$ can be deduced from a crossover between two regimes. Straightforward calculations lead to%

\begin{equation}
\mathrm{Pe}_{\max }\simeq \lambda \dfrac{2/3+U_{sx}/{U_{0}}}{\left( 1+U_{sx}/{%
U_{0}}\right) ^{2}},
\end{equation}

and

\begin{equation}
\sigma _{\max }\simeq \lambda \dfrac{U_{sy}/{U_{0}}}{\left( 2/3+U_{sx}/{U_{0}}%
\right) \left( 1+U_{sx}/{U_{0}}\right) }.
\end{equation}%

Therefore, we can conclude that both $\mathrm{Pe}_{\max }$ and $\sigma _{\max }$ decay when $\phi \rightarrow 1$, and the maximal dispersion is attained at smaller $\phi$.

\bibliographystyle{apsrev4-1}
\bibliography{crossed}

\begin{thebibliography}{33}%
\makeatletter
\providecommand \@ifxundefined [1]{%
 \@ifx{#1\undefined}
}%
\providecommand \@ifnum [1]{%
 \ifnum #1\expandafter \@firstoftwo
 \else \expandafter \@secondoftwo
 \fi
}%
\providecommand \@ifx [1]{%
 \ifx #1\expandafter \@firstoftwo
 \else \expandafter \@secondoftwo
 \fi
}%
\providecommand \natexlab [1]{#1}%
\providecommand \enquote  [1]{``#1''}%
\providecommand \bibnamefont  [1]{#1}%
\providecommand \bibfnamefont [1]{#1}%
\providecommand \citenamefont [1]{#1}%
\providecommand \href@noop [0]{\@secondoftwo}%
\providecommand \href [0]{\begingroup \@sanitize@url \@href}%
\providecommand \@href[1]{\@@startlink{#1}\@@href}%
\providecommand \@@href[1]{\endgroup#1\@@endlink}%
\providecommand \@sanitize@url [0]{\catcode `\\12\catcode `\$12\catcode
  `\&12\catcode `\#12\catcode `\^12\catcode `\_12\catcode `\%12\relax}%
\providecommand \@@startlink[1]{}%
\providecommand \@@endlink[0]{}%
\providecommand \url  [0]{\begingroup\@sanitize@url \@url }%
\providecommand \@url [1]{\endgroup\@href {#1}{\urlprefix }}%
\providecommand \urlprefix  [0]{URL }%
\providecommand \Eprint [0]{\href }%
\providecommand \doibase [0]{http://dx.doi.org/}%
\providecommand \selectlanguage [0]{\@gobble}%
\providecommand \bibinfo  [0]{\@secondoftwo}%
\providecommand \bibfield  [0]{\@secondoftwo}%
\providecommand \translation [1]{[#1]}%
\providecommand \BibitemOpen [0]{}%
\providecommand \bibitemStop [0]{}%
\providecommand \bibitemNoStop [0]{.\EOS\space}%
\providecommand \EOS [0]{\spacefactor3000\relax}%
\providecommand \BibitemShut  [1]{\csname bibitem#1\endcsname}%
\let\auto@bib@innerbib\@empty
\bibitem [{\citenamefont {Quere}(2005)}]{quere.d:2005}%
  \BibitemOpen
  \bibfield  {author} {\bibinfo {author} {\bibfnamefont {D.}~\bibnamefont
  {Quere}},\ }\href@noop {} {\bibfield  {journal} {\bibinfo  {journal} {Rep.
  Prog. Phys.}\ }\textbf {\bibinfo {volume} {68}},\ \bibinfo {pages} {2495}
  (\bibinfo {year} {2005})}\BibitemShut {NoStop}%
\bibitem [{\citenamefont {Darmanin}\ and\ \citenamefont
  {Guittard}(2014)}]{darmanin.t:2014}%
  \BibitemOpen
  \bibfield  {author} {\bibinfo {author} {\bibfnamefont {T.}~\bibnamefont
  {Darmanin}}\ and\ \bibinfo {author} {\bibfnamefont {F.}~\bibnamefont
  {Guittard}},\ }\href@noop {} {\bibfield  {journal} {\bibinfo  {journal} {J.
  Mater. Chem. A}\ }\textbf {\bibinfo {volume} {2}},\ \bibinfo {pages} {16319}
  (\bibinfo {year} {2014})}\BibitemShut {NoStop}%
\bibitem [{\citenamefont {{Bocquet}}\ and\ \citenamefont
  {Barrat}(2007)}]{bocquet2007}%
  \BibitemOpen
  \bibfield  {author} {\bibinfo {author} {\bibfnamefont {L.}~\bibnamefont
  {{Bocquet}}}\ and\ \bibinfo {author} {\bibfnamefont {J.~L.}\ \bibnamefont
  {Barrat}},\ }\href {\doibase 10.1039/B616490K} {\bibfield  {journal}
  {\bibinfo  {journal} {Soft Matter}\ }\textbf {\bibinfo {volume} {3}},\
  \bibinfo {pages} {685} (\bibinfo {year} {2007})}\BibitemShut {NoStop}%
\bibitem [{\citenamefont {Ybert}\ \emph {et~al.}(2007)\citenamefont {Ybert},
  \citenamefont {Barentin}, \citenamefont {Cottin-Bizonne}, \citenamefont
  {Joseph},\ and\ \citenamefont {Bocquet}}]{Ybert2007}%
  \BibitemOpen
  \bibfield  {author} {\bibinfo {author} {\bibfnamefont {C.}~\bibnamefont
  {Ybert}}, \bibinfo {author} {\bibfnamefont {C.}~\bibnamefont {Barentin}},
  \bibinfo {author} {\bibfnamefont {C.}~\bibnamefont {Cottin-Bizonne}},
  \bibinfo {author} {\bibfnamefont {P.}~\bibnamefont {Joseph}}, \ and\ \bibinfo
  {author} {\bibfnamefont {L.}~\bibnamefont {Bocquet}},\ }\href@noop {}
  {\bibfield  {journal} {\bibinfo  {journal} {Phys. Fluids}\ }\textbf {\bibinfo
  {volume} {19}},\ \bibinfo {pages} {123601} (\bibinfo {year}
  {2007})}\BibitemShut {NoStop}%
\bibitem [{\citenamefont {Rothstein}(2010)}]{rothstein.jp:2010}%
  \BibitemOpen
  \bibfield  {author} {\bibinfo {author} {\bibfnamefont {J.~P.}\ \bibnamefont
  {Rothstein}},\ }\href@noop {} {\bibfield  {journal} {\bibinfo  {journal}
  {Annu. Rev. Fluid Mech.}\ }\textbf {\bibinfo {volume} {42}},\ \bibinfo
  {pages} {89} (\bibinfo {year} {2010})}\BibitemShut {NoStop}%
\bibitem [{\citenamefont {{Vinogradova}}\ and\ \citenamefont
  {{Dubov}}(2012)}]{vinogradova.oi:2012}%
  \BibitemOpen
  \bibfield  {author} {\bibinfo {author} {\bibfnamefont {O.~I.}\ \bibnamefont
  {{Vinogradova}}}\ and\ \bibinfo {author} {\bibfnamefont {A.~L.}\ \bibnamefont
  {{Dubov}}},\ }\href@noop {} {\bibfield  {journal} {\bibinfo  {journal}
  {Mendeleev Commun.}\ }\textbf {\bibinfo {volume} {19}},\ \bibinfo {pages}
  {229} (\bibinfo {year} {2012})}\BibitemShut {NoStop}%
\bibitem [{\citenamefont {Feuillebois}\ \emph {et~al.}(2009)\citenamefont
  {Feuillebois}, \citenamefont {Bazant},\ and\ \citenamefont
  {Vinogradova}}]{feuillebois.f:2009}%
  \BibitemOpen
  \bibfield  {author} {\bibinfo {author} {\bibfnamefont {F.}~\bibnamefont
  {Feuillebois}}, \bibinfo {author} {\bibfnamefont {M.~Z.}\ \bibnamefont
  {Bazant}}, \ and\ \bibinfo {author} {\bibfnamefont {O.~I.}\ \bibnamefont
  {Vinogradova}},\ }\href@noop {} {\bibfield  {journal} {\bibinfo  {journal}
  {Phys. Rev. Lett.}\ }\textbf {\bibinfo {volume} {102}},\ \bibinfo {pages}
  {026001} (\bibinfo {year} {2009})}\BibitemShut {NoStop}%
\bibitem [{\citenamefont {{Stone}}\ \emph {et~al.}(2004)\citenamefont
  {{Stone}}, \citenamefont {{Stroock}},\ and\ \citenamefont
  {{Ajdari}}}]{stone2004}%
  \BibitemOpen
  \bibfield  {author} {\bibinfo {author} {\bibfnamefont {H.~A.}\ \bibnamefont
  {{Stone}}}, \bibinfo {author} {\bibfnamefont {A.~D.}\ \bibnamefont
  {{Stroock}}}, \ and\ \bibinfo {author} {\bibfnamefont {A.}~\bibnamefont
  {{Ajdari}}},\ }\href {\doibase 10.1146/annurev.fluid.36.050802.122124}
  {\bibfield  {journal} {\bibinfo  {journal} {Annu. Rev. Fluid Mech.}\ }\textbf
  {\bibinfo {volume} {36}},\ \bibinfo {pages} {381} (\bibinfo {year}
  {2004})}\BibitemShut {NoStop}%
\bibitem [{\citenamefont {Bazant}\ and\ \citenamefont
  {Vinogradova}(2008)}]{bazant2008tensorial}%
  \BibitemOpen
  \bibfield  {author} {\bibinfo {author} {\bibfnamefont {M.~Z.}\ \bibnamefont
  {Bazant}}\ and\ \bibinfo {author} {\bibfnamefont {O.~I.}\ \bibnamefont
  {Vinogradova}},\ }\href@noop {} {\bibfield  {journal} {\bibinfo  {journal}
  {J. Fluid Mech.}\ }\textbf {\bibinfo {volume} {613}},\ \bibinfo {pages} {125}
  (\bibinfo {year} {2008})}\BibitemShut {NoStop}%
\bibitem [{\citenamefont {Schmieschek}\ \emph {et~al.}(2012)\citenamefont
  {Schmieschek}, \citenamefont {Belyaev}, \citenamefont {Harting},\ and\
  \citenamefont {Vinogradova}}]{harting.j:2012}%
  \BibitemOpen
  \bibfield  {author} {\bibinfo {author} {\bibfnamefont {S.}~\bibnamefont
  {Schmieschek}}, \bibinfo {author} {\bibfnamefont {A.~V.}\ \bibnamefont
  {Belyaev}}, \bibinfo {author} {\bibfnamefont {J.}~\bibnamefont {Harting}}, \
  and\ \bibinfo {author} {\bibfnamefont {O.~I.}\ \bibnamefont {Vinogradova}},\
  }\href@noop {} {\bibfield  {journal} {\bibinfo  {journal} {Phys. Rev. E}\
  }\textbf {\bibinfo {volume} {85}},\ \bibinfo {pages} {016324} (\bibinfo
  {year} {2012})}\BibitemShut {NoStop}%
\bibitem [{\citenamefont {Feuillebois}\ \emph {et~al.}(2010)\citenamefont
  {Feuillebois}, \citenamefont {Bazant},\ and\ \citenamefont
  {Vinogradova}}]{mixer2010}%
  \BibitemOpen
  \bibfield  {author} {\bibinfo {author} {\bibfnamefont {F.}~\bibnamefont
  {Feuillebois}}, \bibinfo {author} {\bibfnamefont {M.~Z.}\ \bibnamefont
  {Bazant}}, \ and\ \bibinfo {author} {\bibfnamefont {O.~I.}\ \bibnamefont
  {Vinogradova}},\ }\href@noop {} {\bibfield  {journal} {\bibinfo  {journal}
  {Phys. Rev. E}\ }\textbf {\bibinfo {volume} {82}},\ \bibinfo {pages}
  {055301(R)} (\bibinfo {year} {2010})}\BibitemShut {NoStop}%
\bibitem [{\citenamefont {Vinogradova}\ and\ \citenamefont
  {Belyaev}(2011)}]{vinogradova.oi:2010}%
  \BibitemOpen
  \bibfield  {author} {\bibinfo {author} {\bibfnamefont {O.~I.}\ \bibnamefont
  {Vinogradova}}\ and\ \bibinfo {author} {\bibfnamefont {A.~V.}\ \bibnamefont
  {Belyaev}},\ }\href@noop {} {\bibfield  {journal} {\bibinfo  {journal} {J.
  Phys.: Cond. Matter}\ }\textbf {\bibinfo {volume} {23}},\ \bibinfo {pages}
  {184104} (\bibinfo {year} {2011})}\BibitemShut {NoStop}%
\bibitem [{\citenamefont {Ng}\ and\ \citenamefont {Wang}(2010)}]{ng.co:2010b}%
  \BibitemOpen
  \bibfield  {author} {\bibinfo {author} {\bibfnamefont {C.~O.}\ \bibnamefont
  {Ng}}\ and\ \bibinfo {author} {\bibfnamefont {C.~Y.}\ \bibnamefont {Wang}},\
  }\href@noop {} {\bibfield  {journal} {\bibinfo  {journal} {Microfluid
  Nanofluid}\ }\textbf {\bibinfo {volume} {8}},\ \bibinfo {pages} {361}
  (\bibinfo {year} {2010})}\BibitemShut {NoStop}%
\bibitem [{\citenamefont {Nizkaya}\ \emph {et~al.}(2015)\citenamefont
  {Nizkaya}, \citenamefont {Asmolov}, \citenamefont {Zhou}, \citenamefont
  {Schmid},\ and\ \citenamefont {Vinogradova}}]{nizkaya2015flows}%
  \BibitemOpen
  \bibfield  {author} {\bibinfo {author} {\bibfnamefont {T.~V.}\ \bibnamefont
  {Nizkaya}}, \bibinfo {author} {\bibfnamefont {E.~S.}\ \bibnamefont
  {Asmolov}}, \bibinfo {author} {\bibfnamefont {J.}~\bibnamefont {Zhou}},
  \bibinfo {author} {\bibfnamefont {F.}~\bibnamefont {Schmid}}, \ and\ \bibinfo
  {author} {\bibfnamefont {O.~I.}\ \bibnamefont {Vinogradova}},\ }\href@noop {}
  {\bibfield  {journal} {\bibinfo  {journal} {Phys. Rev. E}\ }\textbf {\bibinfo
  {volume} {91}},\ \bibinfo {pages} {033020} (\bibinfo {year}
  {2015})}\BibitemShut {NoStop}%
\bibitem [{\citenamefont {Ng}\ \emph {et~al.}(2010)\citenamefont {Ng},
  \citenamefont {Chu},\ and\ \citenamefont {Wang}}]{ng.co:2010}%
  \BibitemOpen
  \bibfield  {author} {\bibinfo {author} {\bibfnamefont {C.~O.}\ \bibnamefont
  {Ng}}, \bibinfo {author} {\bibfnamefont {H.~C.~W.}\ \bibnamefont {Chu}}, \
  and\ \bibinfo {author} {\bibfnamefont {C.~Y.}\ \bibnamefont {Wang}},\
  }\href@noop {} {\bibfield  {journal} {\bibinfo  {journal} {Phys. Fluids}\
  }\textbf {\bibinfo {volume} {22}},\ \bibinfo {pages} {102002} (\bibinfo
  {year} {2010})}\BibitemShut {NoStop}%
\bibitem [{\citenamefont {Pimponi}\ \emph {et~al.}(2014)\citenamefont
  {Pimponi}, \citenamefont {Chinappi}, \citenamefont {Gualtieri},\ and\
  \citenamefont {Casciola}}]{pimponi.d:2014}%
  \BibitemOpen
  \bibfield  {author} {\bibinfo {author} {\bibfnamefont {D.}~\bibnamefont
  {Pimponi}}, \bibinfo {author} {\bibfnamefont {M.}~\bibnamefont {Chinappi}},
  \bibinfo {author} {\bibfnamefont {P.}~\bibnamefont {Gualtieri}}, \ and\
  \bibinfo {author} {\bibfnamefont {C.~M.}\ \bibnamefont {Casciola}},\
  }\href@noop {} {\bibfield  {journal} {\bibinfo  {journal} {Microfluidics
  Nanofluidics}\ }\textbf {\bibinfo {volume} {16}},\ \bibinfo {pages} {571}
  (\bibinfo {year} {2014})}\BibitemShut {NoStop}%
\bibitem [{\citenamefont {Asmolov}\ \emph {et~al.}(2015)\citenamefont
  {Asmolov}, \citenamefont {Dubov}, \citenamefont {Nizkaya}, \citenamefont
  {Kuehne},\ and\ \citenamefont {Vinogradova}}]{LabChip}%
  \BibitemOpen
  \bibfield  {author} {\bibinfo {author} {\bibfnamefont {E.~S.}\ \bibnamefont
  {Asmolov}}, \bibinfo {author} {\bibfnamefont {A.~L.}\ \bibnamefont {Dubov}},
  \bibinfo {author} {\bibfnamefont {T.~V.}\ \bibnamefont {Nizkaya}}, \bibinfo
  {author} {\bibfnamefont {A.~J.~C.}\ \bibnamefont {Kuehne}}, \ and\ \bibinfo
  {author} {\bibfnamefont {O.~I.}\ \bibnamefont {Vinogradova}},\ }\href
  {\doibase 10.1039/C5LC00310E} {\bibfield  {journal} {\bibinfo  {journal} {Lab
  Chip}\ }\textbf {\bibinfo {volume} {15}},\ \bibinfo {pages} {2835} (\bibinfo
  {year} {2015})}\BibitemShut {NoStop}%
\bibitem [{\citenamefont {Ou}\ \emph {et~al.}(2007)\citenamefont {Ou},
  \citenamefont {Moss},\ and\ \citenamefont {Rothstein}}]{ou2007enhanced}%
  \BibitemOpen
  \bibfield  {author} {\bibinfo {author} {\bibfnamefont {J.}~\bibnamefont
  {Ou}}, \bibinfo {author} {\bibfnamefont {G.~R.}\ \bibnamefont {Moss}}, \ and\
  \bibinfo {author} {\bibfnamefont {J.~P.}\ \bibnamefont {Rothstein}},\
  }\href@noop {} {\bibfield  {journal} {\bibinfo  {journal} {Phys. Rev. E}\
  }\textbf {\bibinfo {volume} {76}},\ \bibinfo {pages} {016304} (\bibinfo
  {year} {2007})}\BibitemShut {NoStop}%
\bibitem [{\citenamefont {Jaimon}\ and\ \citenamefont
  {Ranjith}(2016)}]{jaimon2016numerical}%
  \BibitemOpen
  \bibfield  {author} {\bibinfo {author} {\bibfnamefont {C.}~\bibnamefont
  {Jaimon}}\ and\ \bibinfo {author} {\bibfnamefont {S.~K.}\ \bibnamefont
  {Ranjith}},\ }\href@noop {} {\bibfield  {journal} {\bibinfo  {journal}
  {Microsyst. Technol.}\ }\textbf {\bibinfo {volume} {24}},\ \bibinfo {pages}
  {1} (\bibinfo {year} {2016})}\BibitemShut {NoStop}%
\bibitem [{\citenamefont {Frey}\ and\ \citenamefont
  {Kroy}(2005)}]{frey.e:2005}%
  \BibitemOpen
  \bibfield  {author} {\bibinfo {author} {\bibfnamefont {E.}~\bibnamefont
  {Frey}}\ and\ \bibinfo {author} {\bibfnamefont {K.}~\bibnamefont {Kroy}},\
  }\href@noop {} {\bibfield  {journal} {\bibinfo  {journal} {Ann. Phys.}\
  }\textbf {\bibinfo {volume} {14}},\ \bibinfo {pages} {20} (\bibinfo {year}
  {2005})}\BibitemShut {NoStop}%
\bibitem [{\citenamefont {Wang}\ \emph {et~al.}(2014)\citenamefont {Wang},
  \citenamefont {Chen}, \citenamefont {Sheng},\ and\ \citenamefont
  {Tsao}}]{wang.z:2014}%
  \BibitemOpen
  \bibfield  {author} {\bibinfo {author} {\bibfnamefont {Z.}~\bibnamefont
  {Wang}}, \bibinfo {author} {\bibfnamefont {H.~Y.}\ \bibnamefont {Chen}},
  \bibinfo {author} {\bibfnamefont {Y.~J.}\ \bibnamefont {Sheng}}, \ and\
  \bibinfo {author} {\bibfnamefont {H.~K.}\ \bibnamefont {Tsao}},\ }\href@noop
  {} {\bibfield  {journal} {\bibinfo  {journal} {Soft Matter}\ }\textbf
  {\bibinfo {volume} {10}},\ \bibinfo {pages} {3209} (\bibinfo {year}
  {2014})}\BibitemShut {NoStop}%
\bibitem [{\citenamefont {Vinogradova}\ \emph {et~al.}(2009)\citenamefont
  {Vinogradova}, \citenamefont {Koynov}, \citenamefont {Best},\ and\
  \citenamefont {Feuillebois}}]{vinogradova.oi:2009}%
  \BibitemOpen
  \bibfield  {author} {\bibinfo {author} {\bibfnamefont {O.~I.}\ \bibnamefont
  {Vinogradova}}, \bibinfo {author} {\bibfnamefont {K.}~\bibnamefont {Koynov}},
  \bibinfo {author} {\bibfnamefont {A.}~\bibnamefont {Best}}, \ and\ \bibinfo
  {author} {\bibfnamefont {F.}~\bibnamefont {Feuillebois}},\ }\href@noop {}
  {\bibfield  {journal} {\bibinfo  {journal} {Phys. Rev. Lett.}\ }\textbf
  {\bibinfo {volume} {102}},\ \bibinfo {pages} {118302} (\bibinfo {year}
  {2009})}\BibitemShut {NoStop}%
\bibitem [{\citenamefont {Ottino}\ and\ \citenamefont
  {Wiggins}(2004)}]{ottino2004introduction}%
  \BibitemOpen
  \bibfield  {author} {\bibinfo {author} {\bibfnamefont {J.~M.}\ \bibnamefont
  {Ottino}}\ and\ \bibinfo {author} {\bibfnamefont {S.}~\bibnamefont
  {Wiggins}},\ }\href@noop {} {\bibfield  {journal} {\bibinfo  {journal}
  {Philos. Trans. A Math. Phys. Eng. Sci.}\ ,\ \bibinfo {pages} {923}}
  (\bibinfo {year} {2004})}\BibitemShut {NoStop}%
\bibitem [{\citenamefont {Stroock}\ \emph {et~al.}(2002)\citenamefont
  {Stroock}, \citenamefont {Dertinger}, \citenamefont {Ajdari}, \citenamefont
  {Mezi{\'c}}, \citenamefont {Stone},\ and\ \citenamefont
  {Whitesides}}]{stroock2002chaotic}%
  \BibitemOpen
  \bibfield  {author} {\bibinfo {author} {\bibfnamefont {A.~D.}\ \bibnamefont
  {Stroock}}, \bibinfo {author} {\bibfnamefont {S.~K.}\ \bibnamefont
  {Dertinger}}, \bibinfo {author} {\bibfnamefont {A.}~\bibnamefont {Ajdari}},
  \bibinfo {author} {\bibfnamefont {I.}~\bibnamefont {Mezi{\'c}}}, \bibinfo
  {author} {\bibfnamefont {H.~A.}\ \bibnamefont {Stone}}, \ and\ \bibinfo
  {author} {\bibfnamefont {G.~M.}\ \bibnamefont {Whitesides}},\ }\href@noop {}
  {\bibfield  {journal} {\bibinfo  {journal} {Science}\ }\textbf {\bibinfo
  {volume} {295}},\ \bibinfo {pages} {647} (\bibinfo {year}
  {2002})}\BibitemShut {NoStop}%
\bibitem [{\citenamefont {Howell~Jr}\ \emph {et~al.}(2008)\citenamefont
  {Howell~Jr}, \citenamefont {Mott}, \citenamefont {Ligler}, \citenamefont
  {Golden}, \citenamefont {Kaplan},\ and\ \citenamefont {Oran}}]{howell2008}%
  \BibitemOpen
  \bibfield  {author} {\bibinfo {author} {\bibfnamefont {P.~B.}\ \bibnamefont
  {Howell~Jr}}, \bibinfo {author} {\bibfnamefont {D.~R.}\ \bibnamefont {Mott}},
  \bibinfo {author} {\bibfnamefont {F.~S.}\ \bibnamefont {Ligler}}, \bibinfo
  {author} {\bibfnamefont {J.~P.}\ \bibnamefont {Golden}}, \bibinfo {author}
  {\bibfnamefont {C.~R.}\ \bibnamefont {Kaplan}}, \ and\ \bibinfo {author}
  {\bibfnamefont {E.~S.}\ \bibnamefont {Oran}},\ }\href@noop {} {\bibfield
  {journal} {\bibinfo  {journal} {J. Micromech. Microeng.}\ }\textbf {\bibinfo
  {volume} {18}},\ \bibinfo {pages} {115019} (\bibinfo {year}
  {2008})}\BibitemShut {NoStop}%
\bibitem [{\citenamefont {Metzler}\ and\ \citenamefont
  {Klafter}(2000)}]{metzler2000}%
  \BibitemOpen
  \bibfield  {author} {\bibinfo {author} {\bibfnamefont {R.}~\bibnamefont
  {Metzler}}\ and\ \bibinfo {author} {\bibfnamefont {J.}~\bibnamefont
  {Klafter}},\ }\href@noop {} {\bibfield  {journal} {\bibinfo  {journal} {Phys.
  Rep.}\ }\textbf {\bibinfo {volume} {339}},\ \bibinfo {pages} {1} (\bibinfo
  {year} {2000})}\BibitemShut {NoStop}%
\bibitem [{\citenamefont {Matheron}\ and\ \citenamefont
  {De~Marsily}(1980)}]{matheron1980}%
  \BibitemOpen
  \bibfield  {author} {\bibinfo {author} {\bibfnamefont {G.}~\bibnamefont
  {Matheron}}\ and\ \bibinfo {author} {\bibfnamefont {G.}~\bibnamefont
  {De~Marsily}},\ }\href@noop {} {\bibfield  {journal} {\bibinfo  {journal}
  {Water Resources Research}\ }\textbf {\bibinfo {volume} {16}},\ \bibinfo
  {pages} {901} (\bibinfo {year} {1980})}\BibitemShut {NoStop}%
\bibitem [{\citenamefont {Krapivsky}\ \emph {et~al.}(2010)\citenamefont
  {Krapivsky}, \citenamefont {Redner},\ and\ \citenamefont
  {Ben-Naim}}]{krapivsky2010}%
  \BibitemOpen
  \bibfield  {author} {\bibinfo {author} {\bibfnamefont {P.~L.}\ \bibnamefont
  {Krapivsky}}, \bibinfo {author} {\bibfnamefont {S.}~\bibnamefont {Redner}}, \
  and\ \bibinfo {author} {\bibfnamefont {E.}~\bibnamefont {Ben-Naim}},\
  }\href@noop {} {\emph {\bibinfo {title} {A kinetic view of statistical
  physics}}}\ (\bibinfo  {publisher} {Cambridge University Press},\ \bibinfo
  {year} {2010})\BibitemShut {NoStop}%
\bibitem [{\citenamefont {Shlesinger}\ \emph {et~al.}(1987)\citenamefont
  {Shlesinger}, \citenamefont {West},\ and\ \citenamefont
  {Klafter}}]{shlesinger1987}%
  \BibitemOpen
  \bibfield  {author} {\bibinfo {author} {\bibfnamefont {M.}~\bibnamefont
  {Shlesinger}}, \bibinfo {author} {\bibfnamefont {B.}~\bibnamefont {West}}, \
  and\ \bibinfo {author} {\bibfnamefont {J.}~\bibnamefont {Klafter}},\
  }\href@noop {} {\bibfield  {journal} {\bibinfo  {journal} {Phys. Rev. Lett.}\
  }\textbf {\bibinfo {volume} {58}},\ \bibinfo {pages} {1100} (\bibinfo {year}
  {1987})}\BibitemShut {NoStop}%
\bibitem [{\citenamefont {Foister}\ and\ \citenamefont {Van
  De~Ven}(1980)}]{foister1980}%
  \BibitemOpen
  \bibfield  {author} {\bibinfo {author} {\bibfnamefont {R.~T.}\ \bibnamefont
  {Foister}}\ and\ \bibinfo {author} {\bibfnamefont {T.~G.~M.}\ \bibnamefont
  {Van De~Ven}},\ }\href {\doibase 10.1017/S0022112080002042} {\bibfield
  {journal} {\bibinfo  {journal} {J. Fluid Mech.}\ }\textbf {\bibinfo {volume}
  {96}},\ \bibinfo {pages} {105} (\bibinfo {year} {1980})}\BibitemShut
  {NoStop}%
\bibitem [{\citenamefont {Dentz}\ and\ \citenamefont
  {Berkowitz}(2003)}]{dentz2003}%
  \BibitemOpen
  \bibfield  {author} {\bibinfo {author} {\bibfnamefont {M.}~\bibnamefont
  {Dentz}}\ and\ \bibinfo {author} {\bibfnamefont {B.}~\bibnamefont
  {Berkowitz}},\ }\href@noop {} {\bibfield  {journal} {\bibinfo  {journal}
  {Water Resources Research}\ }\textbf {\bibinfo {volume} {39}} (\bibinfo
  {year} {2003})}\BibitemShut {NoStop}%
\bibitem [{\citenamefont {Benichou}\ \emph {et~al.}(2013)\citenamefont
  {Benichou}, \citenamefont {Bodrova}, \citenamefont {Chakraborty},
  \citenamefont {Illien}, \citenamefont {Law}, \citenamefont {Mejia-Monasteri},
  \citenamefont {Oshanin},\ and\ \citenamefont {Voituriez}}]{Benichou.o:2013}%
  \BibitemOpen
  \bibfield  {author} {\bibinfo {author} {\bibfnamefont {O.}~\bibnamefont
  {Benichou}}, \bibinfo {author} {\bibfnamefont {A.}~\bibnamefont {Bodrova}},
  \bibinfo {author} {\bibfnamefont {D.}~\bibnamefont {Chakraborty}}, \bibinfo
  {author} {\bibfnamefont {P.}~\bibnamefont {Illien}}, \bibinfo {author}
  {\bibfnamefont {A.}~\bibnamefont {Law}}, \bibinfo {author} {\bibfnamefont
  {C.}~\bibnamefont {Mejia-Monasteri}}, \bibinfo {author} {\bibfnamefont
  {G.}~\bibnamefont {Oshanin}}, \ and\ \bibinfo {author} {\bibfnamefont
  {R.}~\bibnamefont {Voituriez}},\ }\href@noop {} {\bibfield  {journal}
  {\bibinfo  {journal} {Phys. Rev. Lett.}\ }\textbf {\bibinfo {volume} {111}},\
  \bibinfo {pages} {260601} (\bibinfo {year} {2013})}\BibitemShut {NoStop}%
\bibitem [{\citenamefont {Nizkaya}\ \emph {et~al.}(2013)\citenamefont
  {Nizkaya}, \citenamefont {Asmolov},\ and\ \citenamefont
  {Vinogradova}}]{nizkaya2013flow}%
  \BibitemOpen
  \bibfield  {author} {\bibinfo {author} {\bibfnamefont {T.~V.}\ \bibnamefont
  {Nizkaya}}, \bibinfo {author} {\bibfnamefont {E.~S.}\ \bibnamefont
  {Asmolov}}, \ and\ \bibinfo {author} {\bibfnamefont {O.~I.}\ \bibnamefont
  {Vinogradova}},\ }\href@noop {} {\bibfield  {journal} {\bibinfo  {journal}
  {Soft Matter}\ }\textbf {\bibinfo {volume} {9}},\ \bibinfo {pages} {11671}
  (\bibinfo {year} {2013})}\BibitemShut {NoStop}%
\end{thebibliography}%

\end{document}